\documentstyle[12pt]{article}
\normalbaselines
\begin{document}
\bibliographystyle{unsrt}
\renewcommand{\abstractname}{  }\abstractname
\renewcommand{\refname}{  }\refname

\begin{center}
{\bf INERTIAL PARAMETERS AND SUPERFLUID-TO-NORMAL\\
PHASE TRANSITION IN SUPERDEFORMED BANDS}\\
\vspace{2mm}
I. M. Pavlichenkov \\
Russian Research Center "Kurchatov
Institute", Moscow, 123182, Russia\\
\end{center}

\begin{abstract}
The quasiclassically exact solution for the second inertial parameter
$\cal B$ is found in a self-consistent way. It is shown that
superdeformation and nonuniform pairing arising from the rotation
induced pair density significantly reduce this inertial parameter. The new
signature for the transition from pairing to normal phase is suggested in
terms of the variation ${\cal B}/{\cal A}$ versus spin. Experimental data
indicate the existence of such transition in the three SD mass regions.
\end{abstract}


One of the amazing features of superdeformed (SD) rotational bands is
the extreme regularity of their rotational spectra: a SD nucleus is the
best quantum rotor known in nature. In spite of the fact that numerous
theoretical calculations successfully reproduce the measured intraband
$\gamma$-ray energies (see e.g. [1--4])
the underlying microscopic mechanism of this phenomenon is still not well
understood. To explain the SD rotational spectrum regularity we use in
the present paper the parameterization of its energy by the three term
formula
\begin{equation}
E(I)= E_0 + {\cal A}I(I+1) + {\cal B}I^2(I+1)^2,
\label{exp}
\end{equation}
which is valid for an axially symmetric deformed nucleus with $K=0$.
The inertial parameters ${\cal A}=\hbar^2/2\Im^{(1)}$ ($\Im^{(1)}$ is
the kinematic moment of inertia) and ${\cal B}$ are the objects of our
investigation. They are determined by the transition energies
$E_{\gamma}(I)=E(I+2)-E(I)$ as follows
$$
{\cal A}(I)\!=\!\frac{1}{4(2I\!+\!5)}\!\!\left[\!\frac{I^2\!+
\!7I\!+\!13}{2I+3}\!E_{\gamma}(I)\! -\! \frac{I^2\!+\!3I\!+
  \!3}{2I+7}E_{\gamma}(I\!+\!2)\!\right]\!,
$$
\begin{equation}
{\cal B}(I) = \frac{1}{8(2I+5)}\left[ \frac{E_{\gamma}(I+2)}{2I+7}-
\frac{E_{\gamma}(I)}{2I+3}\right].
\label{para}
\end{equation}
The coefficient ${\cal B}(I)$ characterizes the nonadiabatic properties
of a  band and is very sensitive to its internal structure. In particular,
it realizes the relationship of kinematic and dynamic ($\Im^{(2)}$)
moments of inertia. Using the well known expressions for these values
\cite{Fir} and the formulas (\ref{para}) we get
\begin{equation}
{\cal B} = \frac{\hbar^2}{2(2I+3)(2I+7)}\left[ \frac{1}{\Im^{(2)}}-
\frac{2I}{(2I+5)\Im^{(1)}}\right].
\end{equation}
The ratio ${\cal B}/{\cal A}$ determines the convergence radius
\cite{B/M} of the two parameter formula (\ref{exp}), which is of order
100 for the bands in the 80 and 150 mass regions and 40 in the 130 and
190 ones. Faster convergence is obtained with Harris formula
\begin{equation}
E(\omega)= E_0 + \frac{1}{2}\alpha\omega^2 + \frac{3}{4}\beta\omega^4,
\label{Hexp}
\end{equation}
which is based on the fourth-order cranking expansion
\begin{equation}
\alpha = \frac{1}{\omega}Sp(\ell_x\rho^{(1)}),  \hspace{5mm}
 \beta = \frac{1}{\omega^3}Sp(\ell_x\rho^{(3)}),
\label{cranc}
\end{equation}
where $\rho^{(n)}$ is the $n$th correction to the nucleus density matrix,
$\ell_x$ is the single-particle angular momentum projection on the
rotational axis $x$ perpendicular to the symmetry axis $z$, and $\omega$
is the rotational frequency. It follows from Eqs.\ (\ref{exp}) and
(\ref{Hexp}) that
\begin{equation}
\alpha=\hbar^2/2{\cal A},\hspace{5mm} \beta=-\hbar^4{\cal B}/4{\cal A}^4.
\label{cone}
\end{equation}
For simplicity we will deal with the parameter $\beta$.

The problem of the microscopic calculation of the parameter $\cal B$ for
normal deformed (ND) nuclei has attracted considerable attention. Its
value is formed mainly by four effects: vibration-rotation interaction,
centrifugal stretching, perturbation of the quasiparticle motion, and
attenuation of pair correlation by the Coriolis force (Mottelson-Valatin
effect). As it has already been shown in the first attempts at obtaining
${\cal B}$ \cite{GP,Mar}, the latter two effects are dominant for well
deformed nuclei. Unfortunately the results of these and many other works
cannot be used for the superdeformation. The formulas of Ref. \cite{Mar}
have been obtained in the limit of the monopole pairing interaction (the
uniform pairing), which is not adequate at SD shapes as shown in Ref.
\cite{HN}. In the more sophisticated work \cite{GP} the gauge invariant
pairing interaction allows to study the effect of nonuniform pairing.
However this approach is also limited because it neglects the coupling
between major shells. Thus, despite a number of publications on the
subject the correct cranking selfconsistent solution for the
${\cal B}$ coefficient has not been found.

We used the quasiclassical method of Ref. \cite{GP} to derive the following
expression for the $\beta$ parameter in the superfluid phase
\begin{equation}
\beta_s\! =\!-\frac{\hbar^4}{4\Delta^2}\! \sum\!\ell^x_{12}\ell^x_{23}
    \ell^x_{34}\ell^x_{41}F(x_{12},x_{23},x_{34},x_{41})
     \delta(\varepsilon_1\!\!-\!\varepsilon_F\!),
\label{spur}
\end{equation}
where the summation indices $i$=1,2,3,4 refer to the single-particle states
$i$ of the nonrotating mean field with the energy $\varepsilon_i$. The
$\delta$-function means that the summation over the states 1 is restricted
by a small interval at the Fermi energy $\varepsilon_F$ \cite{Mig}. The
dimensionless values $x_{ii'}=(\varepsilon_i\!-\!\varepsilon_{i'})/2\Delta$,
where $\Delta$ is the state independent pairing gap at $\omega=0$,
correspond to energy differences between states permitted by the selection
rules for the matrix element of $\ell_ x$. The function $F$ depending on
these values may be written as follows
\begin{equation}
F\!=\!\!\sum^3_{k=0}\!\hat P_kG_{12}+
   \!\sum^1_{k=0}\!\hat P_kH_{13}
      \!-\!8\!D^2_2 x_{12}x_{23}x_{34}x_{41}h(x_{13}),
\label{fun}
\end{equation}
where the permutation operators in the space of four 
indices $i$, $i\ mod4=i$,
\begin{equation}
\hat P_kx_{i,i'}=x_{i+k,i'+k},
\end{equation}
are used to simlify the formulas. The expressions for $G_{12}$ 
and $H_{13}$ involve the well known functions \cite{Mig}
\begin{equation}
h(x)= (1+x^2)g(x), \hspace{5mm} g(x)=\frac{argshx}{x\sqrt{1+x^2}},
\end{equation}
and have the form:
$$
G_{12}\!=\!\frac{g(x_{12})}{x_{23}x_{41}x_{13}x_{24}}
   \!\left\{\!(\!1\!-\!D_1 x^2_{12})\! \left[\!-1\!\!-\!x^2_{12}\!-
      \!x_{23}x_{41}\!+\!D_1[x^2_{23}(\!1\!-\!x_{12}x_{24})\!+
      \!x^2_{41}(\!1\!+\!x_{12}x_{13})]\!\right.\right.
$$
$$
  +\left.\left.\!D^2_1x_{23}x_{34}x_{41}(x_{23}\!+\!x_{41})
               D^3_1x_{12}x^2_{23}x_{34}x^2_{41}\right] +
    D_1(x_{34}-D_1x_{12}x_{23}x_{41})(x_{34}\right.
$$
$$
        +\left.x_{12}x_{13}x_{24}-D_1x_{12}x_{23}x_{41})\right\},
$$
\begin{equation}
\hspace{3pt}
H_{13}\!=\!\frac{h(x_{13})}{x_{12}x_{23}x_{34}x_{41}}\!\left[\!1\!-
\!D_1( x^2_{12}\!+\!x^2_{23}\!+\!x^2_{34}\!+\!x^2_{41})^2
+D^2_1( x_{12}x_{41}\!+\!x_{23}x_{34})\right],
\end{equation}

\vspace{2mm}
Eq.\ (\ref{spur}) multiplied by $\omega^3$ is the third order cranking
correction to the total angular momentum of the neutron or proton
system. Its derivation will be described in a forthcoming paper. We
want now to emphasize that Eq.\ (\ref{spur}) represents the first
theoretically correct expression for the high order effect of the
Coriolis-pairing interaction at fixed deformation. The result is
obtained by taking into account the effect of rotation on the
Cooper pairs in the gauge invariant form. This effect is described by
the first and the second corrections to the pairing energy
\begin{equation}
\Delta^{(1)}({\bf r})= -\frac{i\hbar^2\omega}{2\Delta}D_1\dot\ell_x,
\hspace{5mm}
\Delta^{(2)}({\bf r})= \frac{\hbar^4\omega^2}{4\Delta^3}D_2\dot\ell^2_x,
\label{nuif}
\end{equation}
where $D_1$ and $D_2$ are the amplitudes of the nonuniform pairing
fields, which are found in a self-consistent way. Let us note that
$\Delta^{(n)}$ is the function of the space coordinates because
$\hbar\dot\ell_x=y\partial U/\partial z-z\partial U/\partial y$,
where $U$ is the potential of a mean field. It is seen that for
the oscillator potential $\dot\ell_x\sim yz$ (the $Y_{21}$ pairing).
The coordinate dependent pairing field is crucial for conservation
of a nucleon current. The theory incorporating the nonuniform pairing
allows also to consider the different limiting cases for the inertial
parameters, which make possible the study of an interplay between
rotation, pairing correlations, and mean field deformation in a SD
band.

In order to consider this problem quantitatively we will use the
axially deformed  oscillator potential with the frequencies 
$\omega_x=\omega_y$ and $\omega_z$ on 
the corresponding axes. In this model the matrix
element $\ell^x_{12}$ is non-zero for the two types of transitions:\\
({\it i}) transitions inside a single oscillator shell (close transitions),
for which $x_{12}\!=\!\pm\nu_1$; ({\it ii}) transitions over a shell
(distant transitions) with $x_{12}\!=\!\pm\nu_2$. The quantities
$\nu_1$ and $\nu_2$ are the well known parameters involved in
the moment of inertia \cite{Mig}:
\begin{equation}
\nu_{1,2}\!=\!\frac{\hbar(\omega_x\!\mp\!\omega_z)}{2\Delta}\!=
\!\frac{k\mp1}{2\xi k^{2/3}},\hspace{5mm}
 \xi\!=\!\frac{\Delta}{\hbar\omega_\circ},
\label{nu}
\end{equation}
where $\hbar\omega_\circ\!=\!41A^{-1/3}$\,MeV. Here and later we use
the axis or frequency ratio $k=c/a=\omega_x/\omega_z$ and the volume
conservation condition. Both of the values $\nu_1$ and $\nu_2$ are large
for superdeformation.
For the fixed state 1, there are 36 different combinations of these base
transitions in the sum of Eq.\ (\ref{spur}). The summation over the states
1 is performed in the Thomas-Fermi approximation. The final expression
for the parameter $\beta_s$ in the oscillator potential is
\begin{equation}
\beta_s = \frac{(k+1)^4}{1875\hbar^2k^{4/3}}AM^3R^6\Phi(\xi,k),
\label{spur1}
\end{equation}
where $R=1.2A^{1/3}$\,fm is the radius of the sphere, which volume is
equal to that of the spheroid with the half-axes $a<c$, $M$ is the nucleon
mass, and A is the number of nucleons. The function $\Phi$ along with its
limiting cases is shown in Fig.\ 1. It is seen that nonuniform
pairing reduces substantially the parameter $\beta_s$ in agreement
with the estimation of Ref. \cite{HN}. On the other hand, the contribution
of distant transitions is minor for small $\xi$. Nevertheless the later are
necessary to obtain the hydrodynamic limit (see below). Since
$\Phi\sim 1$ for a reasonable
pairing gap, $\Delta\sim$0.5\,MeV, the order of the value $\beta_s$ is
$\hbar^4(A/\varepsilon_F)^3$. This, along with the estimation
${\cal A}\sim\varepsilon_FA^{-5/3}$, gives ${\cal B}/{\cal A}\sim A^{-2}$,
which overestimates the minimal value of this ratio in all the SD mass
regions. Thus a small $\Delta$ and nonuniform pairing does not solve
the problem of the SD band regularity.

Let us consider first the limiting case $\Delta=0$. The right part of
Eq.\ (\ref{spur}) vanishes for noncorrelated nucleons. This result is
the artifact of the quasiclassical approximation used in Eq.\
(\ref{spur}). The correct expression for the $\beta$ parameter in the
normal phase, obtained with the limiting values of the
Bogolubov amplitudes ($u_i=0$, $v_i=1$ for $n_i=1$ and $u_i=1$,
$v_i=0$ for $n_i=0$, where $n_i$ is the nucleon occupation numbers),
has the form
\begin{equation}
\beta_n =-\hbar^4 \sum\ell^x_{12}\ell^x_{23}
    \ell^x_{34}\ell^x_{41}\sum^3_{k=0}\!\hat P_k\Bigl\{\frac{n_1}
{\varepsilon_{12}\varepsilon_{13}\varepsilon_{14}}\Bigr\}.
\label{spur2}
\end{equation}
The odd function of the differences
$\varepsilon_{ii'}=\varepsilon_i-\varepsilon_{i'}$ leads to the cancelation
of the main terms in sum (\ref{spur2}) that decreases substantially the
value of $\beta_n$, $\beta_n\sim\hbar^4A^{7/3}/\varepsilon^3_F$. Note
that the centrifugal stretching effect has the same order
$\beta_{str}\sim\beta_n$. Its contribution is small compared to that of
$\beta_s$, but it should not be overlooked for an unpaired system. For
the oscillator potential, we have the following expression
\begin{equation}
\beta_n+\beta_{str} = \frac{k^4-10k^2+1}
                                {6\omega^2_\circ k^{4/3}}AMR^2.
\label{spur3}
\end{equation}
It is seen that $\beta_n\!+\!\beta_{str}\!<\!0$ for the prolate nuclei
with $c/a\!<3.15$, whereas $\beta_s$ is always positive. Thus, with an
increase of the spin $I$, the ratio ${\cal B}/{\cal A}$ has to change sign
and to approach its limiting value ${\cal B}_n/{\cal A}_n\sim A^{-8/3}$,
$\sim 10^{-6}$ for the SD bands in the 130 and 150 mass region, where
rapid rotation destroys pairing correlations. The limiting ratio for a
nucleus consisting of $Z$ protons and $N$ neutrons is expressed by
\begin{equation}
\frac{{\cal B}_n}{{\cal A}_n}\!=\!-2.56\,\frac{(k^4\!-\!10k^2\!+\!1)k^{2/3}}
   {(k^2+1)^3A^{8/3}}\left[\Bigl(\frac{2Z}{A}\!\Bigr)^{1/3}\!+
      \!\Bigl(\frac{2N}{A}\!\Bigr)^{1/3}\right].
\label{ratio}
\end{equation}

One can therefore conclude that there are two distinct regions in the
variation of ${\cal B}/{\cal A}$ versus $I$. The lower part of a SD band
is characterized by a gradual decrease of the pairing gap $\Delta$.
According to Eq.\ (\ref{spur1}) the ratio ${\cal B}/{\cal A}$ should
exhibit a sharp increase. Then it changes sign and approaches the
plateau (\ref{ratio}) at the top of a band because the deformation $c/a$
depends weakly on spin in the normal phase.
Such behavior of the ${\cal B}/{\cal A}$ ratio is the signature of the
pairing phase transition.

We have analyzed all the SD bands of Ref. \cite{Fir} with known or
suggested spins of levels. Figure\ 2 shows the variation of the
${\cal B}/{\cal A}$ ratio with $I$ for bands with different
internal structure and different rotational frequencies. Apart from the
bands $^{192}$Hg(1) and $^{194}$Hg(3), where frequencies are so low
that ${\cal B}/{\cal A}$ rises continuously in the superfluid phase,
and $^{84}$Zr(1), for which pairing is quenched completely and
${\cal B}/{\cal A}$ is close to the limiting value (\ref{ratio}), all
other bands display the behavior described above. It is important to note
that such behavior is observed for  the ND yrast band of $^{84}$Zr,
where ${\cal B}/{\cal A}$  reaches the same limiting value (\ref{ratio})
as in the SD band $^{84}$Zr(1). There are other ND yrast bands of
$^{168}$Yb and $^{168}$Hf with the phase transition which
experimental evidence has been discussed previously in terms of the
canonical variables \cite{Gar} and the spectrum of single-particle
states \cite{Gar,Oliv}.  These bands exhibit the same features. Thus
the plots of Fig.\ 2 demonstrate the universality of the
superfluid-to-normal phase transition for SD and ND bands.
The manifestation of this universality was previously observed in
the anomalous small value ${\cal B}/{\cal A}\simeq 7\times 10^{-6}$
for the ND yrast band of $^{168}$Hf \cite{Chap} and in the weak
dependence of $\Im^{(1)}$ in the upper part of the SD band
$^{152}$Dy(1) \cite{Svia}. The new feature observed in this article
is the small decrease of ${\cal B}/{\cal A}$ at the top of the
$^{152}$Dy(1), $^{132}$Ce(1), $^{84}$Zr(1) and other SD bands. It
may be explained by the decrease of the first multiplier in r.h.s.
of Eq.\ (\ref{ratio}) due to increase of nuclear deformation at
highest spins.

All of the theoretical formulas obtained refer to collective rotation
only. The influence of single-particle degrees of freedom on the
${\cal B}/{\cal A}$ ratio is the subject of a separate investigation.
Preliminary estimations show that the contribution of
an odd nucleon in ${\cal B}_s$ may be positive and
irregular. Visible evidences of the nonrotational degree of freedom are
shown in Fig.\ 3. A point of particular interest is the staggering
band $^{149}$Gd(1). It is seen that the staggering pattern is reproduced
very well and the error bars are smaller than the staggering amplitude. We
would like to emphasize that the physical values adequate for describing
the staggering phenomenon are the ${\cal B}$ parameter or the
${\cal B}/{\cal A}$ ratio. They do not require any smooth reference, which
introduces an ambiguousness in the amplitude and the phase of oscillations.

The next limit we want to consider is that of a large pairing gap $\Delta$.
In this case, the nonuniform pairing (\ref{nuif}) is essential and the
leading terms in the function $\Phi$ are those proportional to the
powers of $D_1$ and $D^2_2$. They result in the limiting
expression $\Phi\sim (\hbar\omega_\circ/\Delta)^2$. Thus for the
very strong pairing ($\Delta\gg\hbar\omega_\circ$), when the size
of the Cooper pair $R\hbar\omega_\circ/\Delta$ becomes much less
than the nuclear radius, the value ${\cal B}_s$ vanishes in agreement
with the hydrodynamic equations of the ideal liquid \cite{B/M}. In the
limit of an extremely large deformation, $c/a\to \infty$, a needle
shaped nucleus with pairing correlations rotates as a rigid body,
$\Im=\Im_{rig}$, ${\cal B}_s=0$. For the finite but large deformation
the deviations from these values are proportional to $(a/c)^{4/3}$. This
means that all nucleons with the exclusion of a small sphere in the
center of a nucleus are completely involved in rotational motion.
Finally, for small deformations we have ${\cal B}_s\sim (c/a-1)^{-6}$
that is comparable to the vibration-rotation interaction \cite{GP}.

Unlike the limiting value (\ref{ratio}), it is impossible to compare with
experiment the ratio ${\cal B}_s/{\cal A}_s$ because the proton
($\Delta_\pi$) and neutron ($\Delta_\nu$) pairing gaps are unknown
for the SD bands. In such a case we try to solve an inverse problem.
The equations for the two inertial parameters
\begin{equation}
\alpha A/\Im_{rig}=Z\varphi(\xi_\pi, k)+N\varphi(\xi_\nu, k),
\label{del1}
\end{equation}
\vspace{-5mm}
\begin{equation}
\hspace{10pt} \beta A/\beta_\circ=Z\Phi(\xi_\pi, k)+N\Phi(\xi_\nu, k),
\label{del2}
\end{equation}
allow in principle to find $\xi_\pi$ and $\xi_\nu$ ($\xi_l=
\Delta_l/\hbar\omega_{\circ l}$, $\omega_{\circ l}=
\omega_{\circ}(2A_l/A)^{2/3}$, $l=\pi, \nu$). Here the function
$\varphi$ is taken from Ref. \cite{Mig},
\begin{equation}
\varphi(\nu_1,\nu_2)=1-\frac{(\nu^2_1-\nu^2_2)^2g(\nu_1)g(\nu_2)}
    {(\nu^2_1+\nu^2_2)[\nu^2_1g(\nu_1)+\nu^2_2g(\nu_2)]},
\end{equation}
and the rigid body moment of inertia and the value $\beta_\circ$ have
the form:
$$
 \Im_{rig}=7.29A^{5/3}(k^2+1)k^{-2/3}10^{-3}\hbar^2MeV^{-1},
$$
\begin{equation}
\hspace{10pt} \beta_\circ=
         2.59A^4(k+1)^4 k^{-4/3}10^{-8}\hbar^4MeV^{-3}.
\label{bet0}
\end{equation}
Finally the axis ratio $k=c/a$ is determined by the quadrupole moment
\begin{equation}
 Q_0=6.05A^{2/3}(k^2-1)k^{-2/3}10^{-3} eb.
\label{qud}
\end{equation}
Unfortunately the system (\ref{del1})--(\ref{del2}) does not have a 
solution. Table\ I gives the solutions of Eqs.\ (\ref{del1}) and (\ref{del2})
under simplifying assumption that the neutron and proton pairing
gaps are equal. Different bands reveal the same tendency.
The obtained values $\Delta_1$ and $\Delta_2$ show that the
oscillator potential overestimates the moment of inertia and, to
a greater extent, the $\beta$ parameter. It is essential to use a
more realistic potential for extracting pairing gaps from the
inertial parameters.

\vspace{5mm}
\small
Table\ I. Pairing gap energies $\Delta_1$ and $\Delta_2$ (in MeV)
found as the independent solutions of Eqs.\ (\ref{del1}) and
(\ref{del2}) correspondingly in the case of $\xi_\pi=\xi_\nu$. The values
$c/a$, $\Im_{rig}$, and $\beta_\circ$ were calculated by using
Eqs.\ (\ref{bet0}) and (\ref{qud}). The experimental values of $\alpha=\Im$
and $\beta$ were taken from Refs.\ \cite{B/M} ($^{172}$Hf(yr)) and
\cite{Bec} ($^{192,4}$Hg(1,2)). Those of $^{236}$U(fi) were determined
from Eqs. (\ref{para}), (\ref{cone}). \\
\vspace{-5mm}
\normalsize
\begin{center}
\begin{tabular}{|c|ccccc|}
\hline
\vspace{-8pt}
  & & & & &\\
 Band&  $c/a$ & $\Im/\Im_{rig}$ & $\Delta_1$
       &  $\beta/\beta_0$ & $\Delta_2$\\ [5pt]
\hline
\vspace{-8pt}
  & & & & &\\
 $^{172}$Hf(yr) & 1.32 & 0.373 & 1.44 & 0.309 & 0.02   \\
 $^{192}$Hg(1) & 1.61 & 0.762 & 0.95 & 0.151 & 0.05  \\
 $^{194}$Hg(1) & 1.60 & 0.793 & 0.83 & 0.112 & 0.04   \\
 $^{236}$U(fi)   & 1.84 & 0.816 & 1.01 & 0.141 & 0.08   \\ [3pt]
\hline
\end{tabular}
\end{center}

\vspace{6mm}
In summary, the exact solution for the inertial parameter $\cal B$ in the
superfluid phase allows to show that neither superdeformation nor
nonuniform pairing arising from rotation induced pair density is
responsible for the regularity of the SD rotational spectra.
The extreme regularity of the SD bands in the 80, 130 and 150 mass
regions is explained by the transition from the superfluid to normal
phase. The new signature of this transition reveals itself in the
characteristic dependence of the ratio ${\cal B}/{\cal A}$ with the
spin I. Application of this criterion to experimental data indicates the
existence of the phase transition in the SD bands of the three SD
mass regions. A new extraction method of the proton and neutron
pairing gap from the $\gamma$-ray energies of interband transitions
is discussed.

\newpage
\normalsize
\begin{center}
FIGURE CAPTIONS
\end{center}

\vspace{3mm}
Fig.\ 1. Plot of the function $\Phi$ from Eq.\ (\ref{spur1}) against the
dimensionless value $\xi$ for the axis ratio
$c/a=2$. The solid, dotted, and dashed lines correspond  respectively
to the exact value, the limit of close transitions, and the uniform
pairing. The scale on the abscissa should be multiplied by a factor of
approximately 7.7 for nuclei in the $A\sim$ 150 mass region to obtain
a gap energy in MeV.

\vspace{3mm}
Fig.\ 2. ${\cal B}/{\cal A}$ ratio versus spin for the SD and ND bands
with mainly collective behavior. The formulas (\ref{para}) are used
to obtain this ratio from the experimental data for the ND (full
circles) and SD (open circles) bands. The solid straight line is the
limiting value ${\cal B}_n/{\cal A}_n$ with the deformation $c/a$
found from the quadrupole moment (\ref{qud}). Error bars (if they are
greater than symbols) include $\gamma$-ray energy uncertainties only.
Uncertainties in a spin assignment are not important for all the SD
bands except $^{152}$Dy(1), as the variation of spins in 2$\hbar$
would merely shift the curves along axis of abscissas.

\vspace{3mm}
Fig.\ 3. Same as Fig.\ 2 for the bands with high-$N$ configurations
having the non-zero alignment $i$. Straight line is the value of
${\cal B}_n/{\cal A}_n$ for $^{153}$Dy(1).

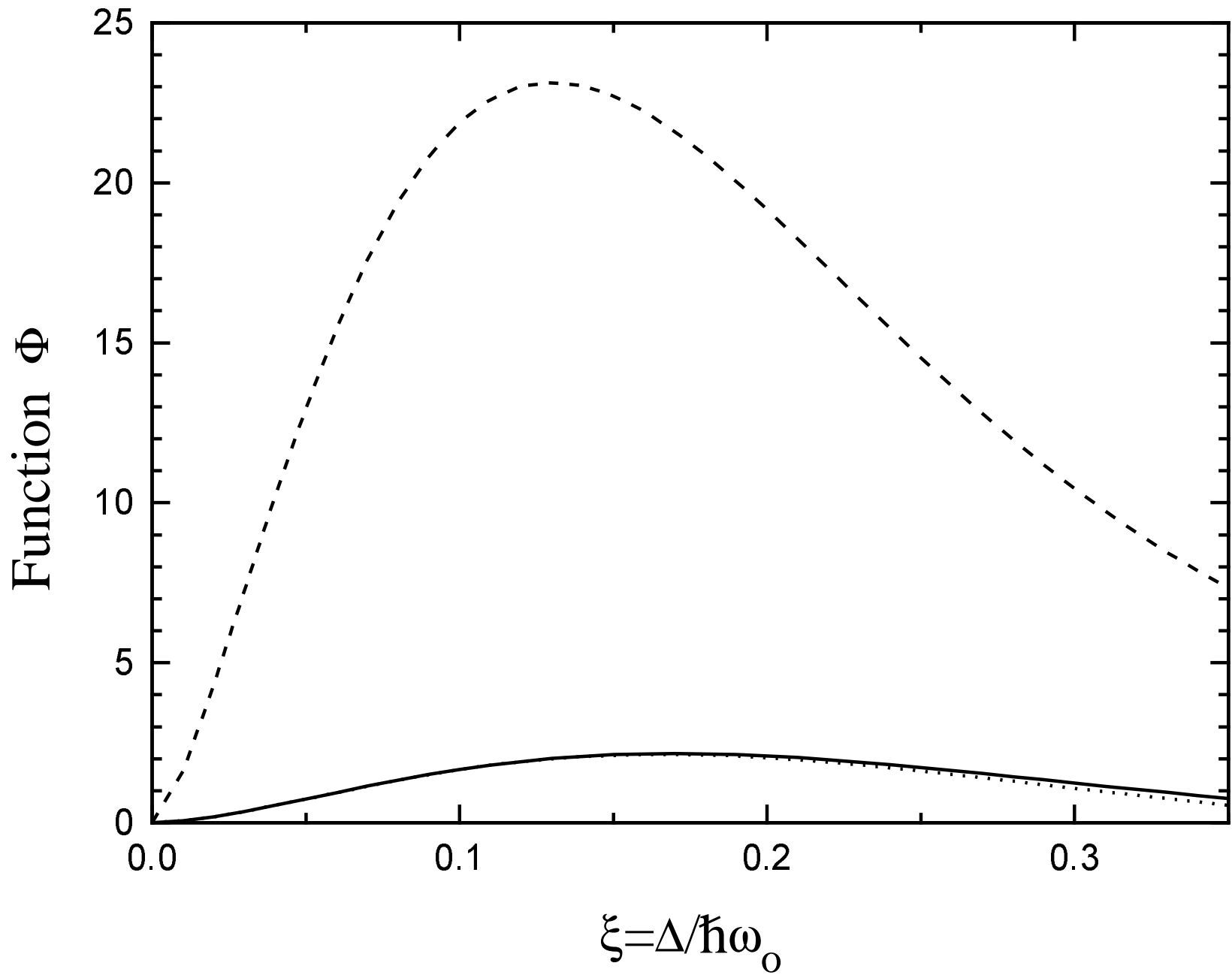
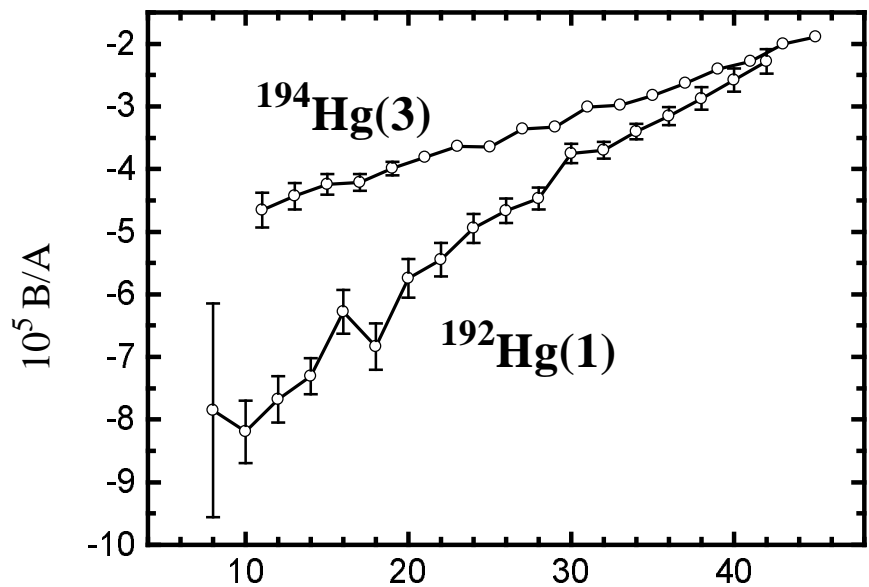
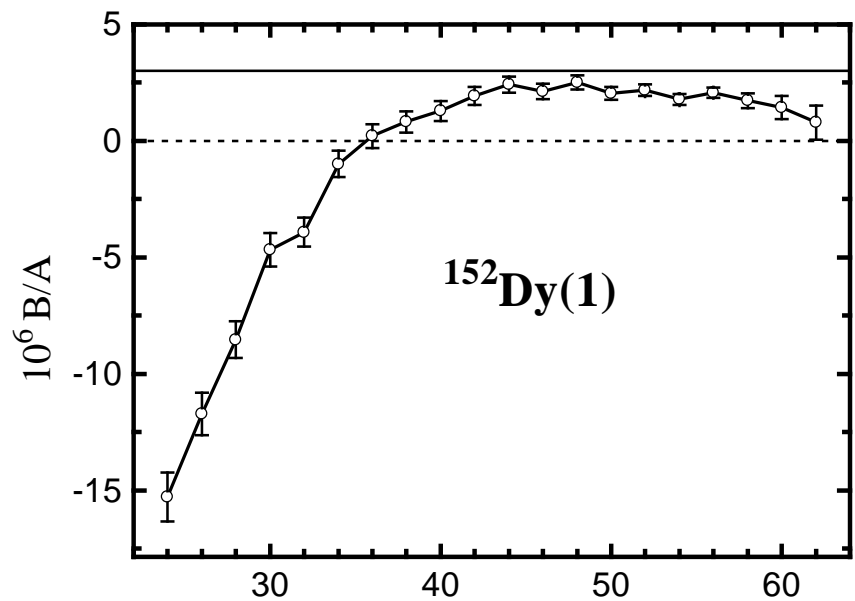
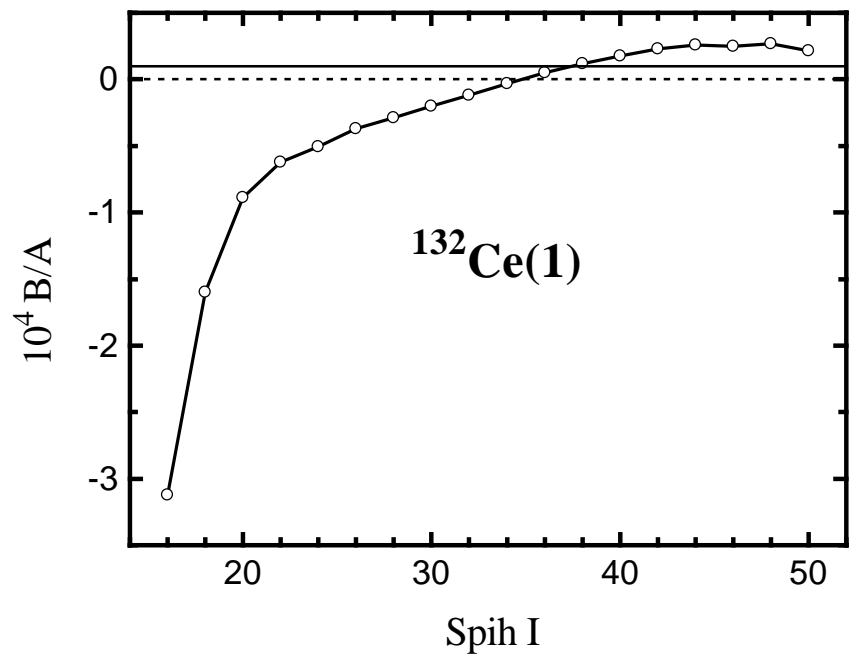
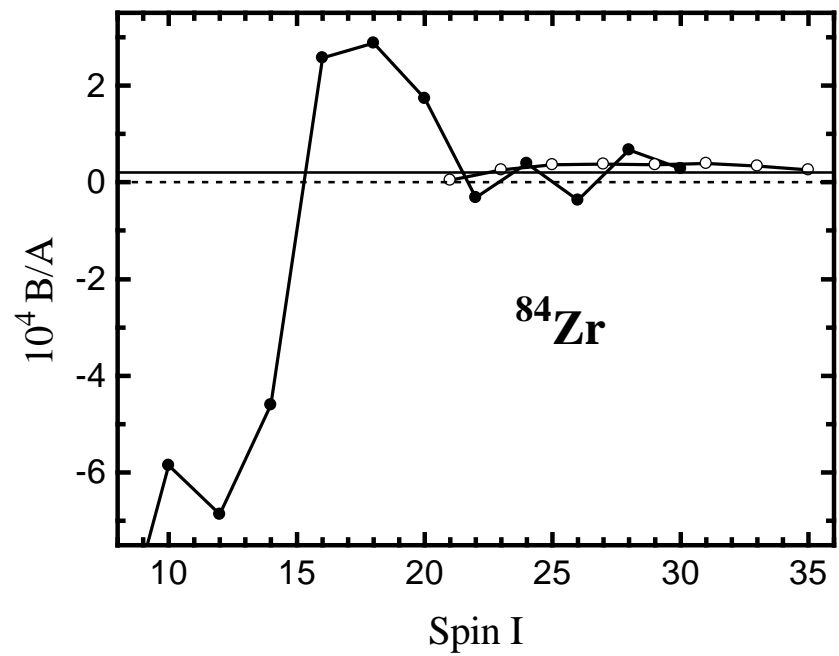
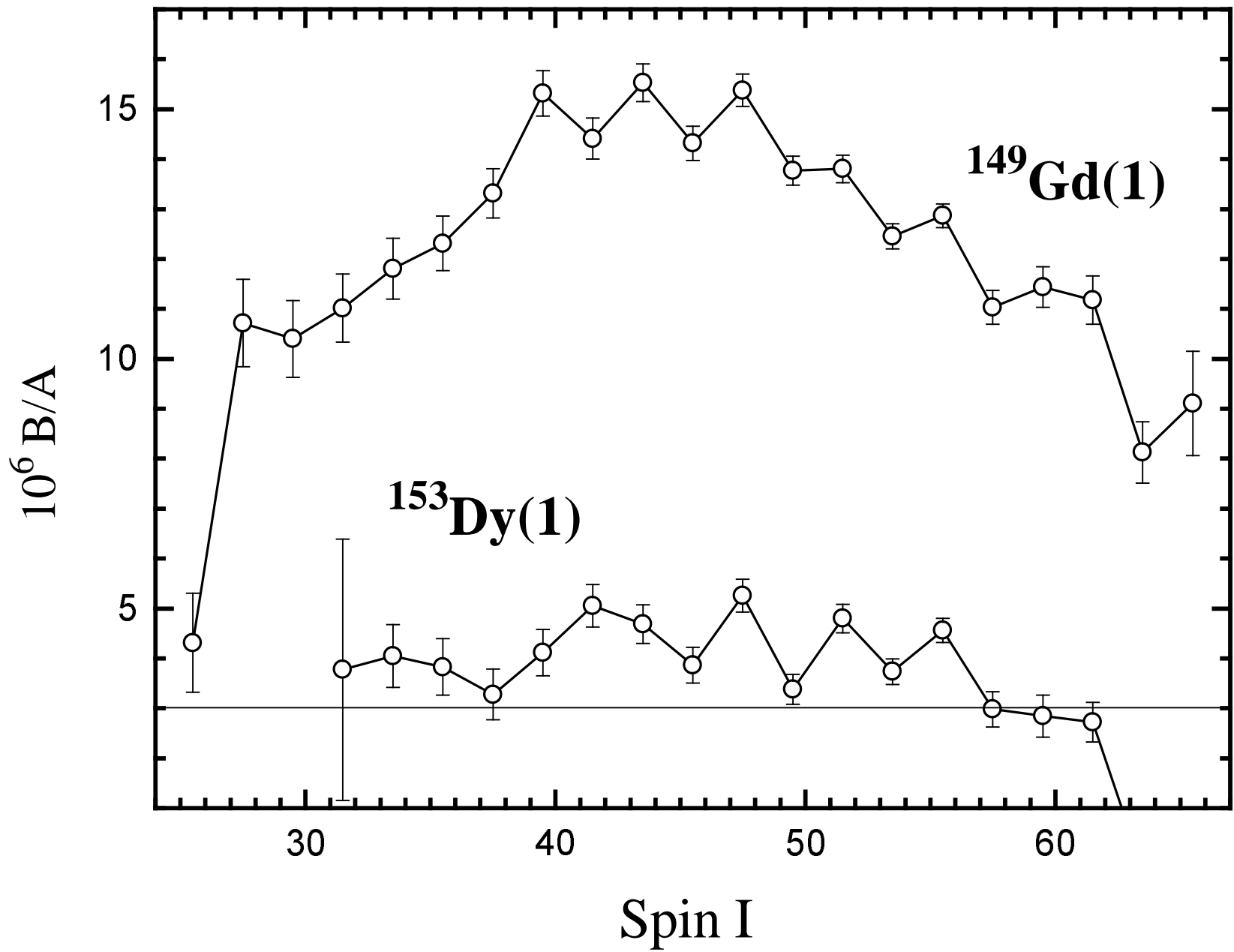
\end{document}